\pdfoutput=1

\documentclass[11pt]{article}

\usepackage{naacl2021}

\usepackage{times}
\usepackage{latexsym}
\usepackage{graphicx}
\usepackage{amsmath}
\usepackage{amssymb}
\usepackage{cprotect}

\usepackage[T1]{fontenc}

\usepackage[utf8]{inputenc}

\usepackage{microtype}

\title{On the Embeddings of Variables \\
in Recurrent Neural Networks for Source Code}

\author{Nadezhda Chirkova \\
  HSE University\thanks{\ \ The work was done while working at Samsung-HSE Laboratory, HSE University.} \\
  Moscow, Russia}

\begin{document}
\maketitle
\begin{abstract}
Source code processing heavily relies on the methods widely used in natural language processing (NLP), but involves specifics that need to be taken into account to achieve higher quality. An example of this specificity is that the semantics of a variable is defined not only by its name but also by the contexts in which the variable occurs. In this work, we develop dynamic embeddings, a recurrent mechanism that adjusts the learned semantics of the variable when it obtains more information about the variable's role in the program. We show that using the proposed dynamic embeddings significantly improves the performance of the recurrent neural network, in code completion and bug fixing tasks.
\end{abstract}

\section{Introduction}
Deep learning is now actively being deployed in source code processing (SCP) for solving such tasks as code completion~\citep{pointer}, generating code comments~\citep{code2seq}, and fixing errors in code~\citep{varmisuse}. Source code visually looks like a text, motivating the wide use of NLP architectures in SCP. A lot of modern SCP approaches are based on recurrent neural networks~\cite{RNNproof}, other popular architectures are transformers, and convolutional and graph neural networks.

Utilizing the specifics of source code as a data domain may potentially improve the quality of neural networks for SCP. These specifics include three main aspects. Firstly, the source code is strictly structured, i.\,e.\, the source code follows the syntactic rules of the programming language. Secondly, the vocabularies may be large or even potentially unlimited, i.\,e.\,a programmer is allowed to define the identifiers of the arbitrary complexity. Thirdly, the identifiers are invariant to renaming, i.\,e.\,renaming all the user-defined identifiers does not change the algorithm that the code snippet implements. The first two mentioned specifics have been extensively investigated in the literature. For example, the tree-based 
architectures, such as TreeLSTM~\citep{tree_lstm} or TreeTransformer~\cite{tree_encoding} allow for the utilization of the code’s structure. On the other hand, using byte-pair encoding~\cite{bpe_code,bpe_nlp} or the anonymization of out-of-vocabulary identifiers~\cite{OOV_anonym} deals with the unlimited vocabulary problem. 
 However, the property of source code being invariant to renaming user-defined identifiers has not been paid much attention to. In this work, we aim to close this gap for the recurrent neural networks (RNNs).

\begin{figure*}[h!]
    \centering
         \includegraphics[width=\linewidth]{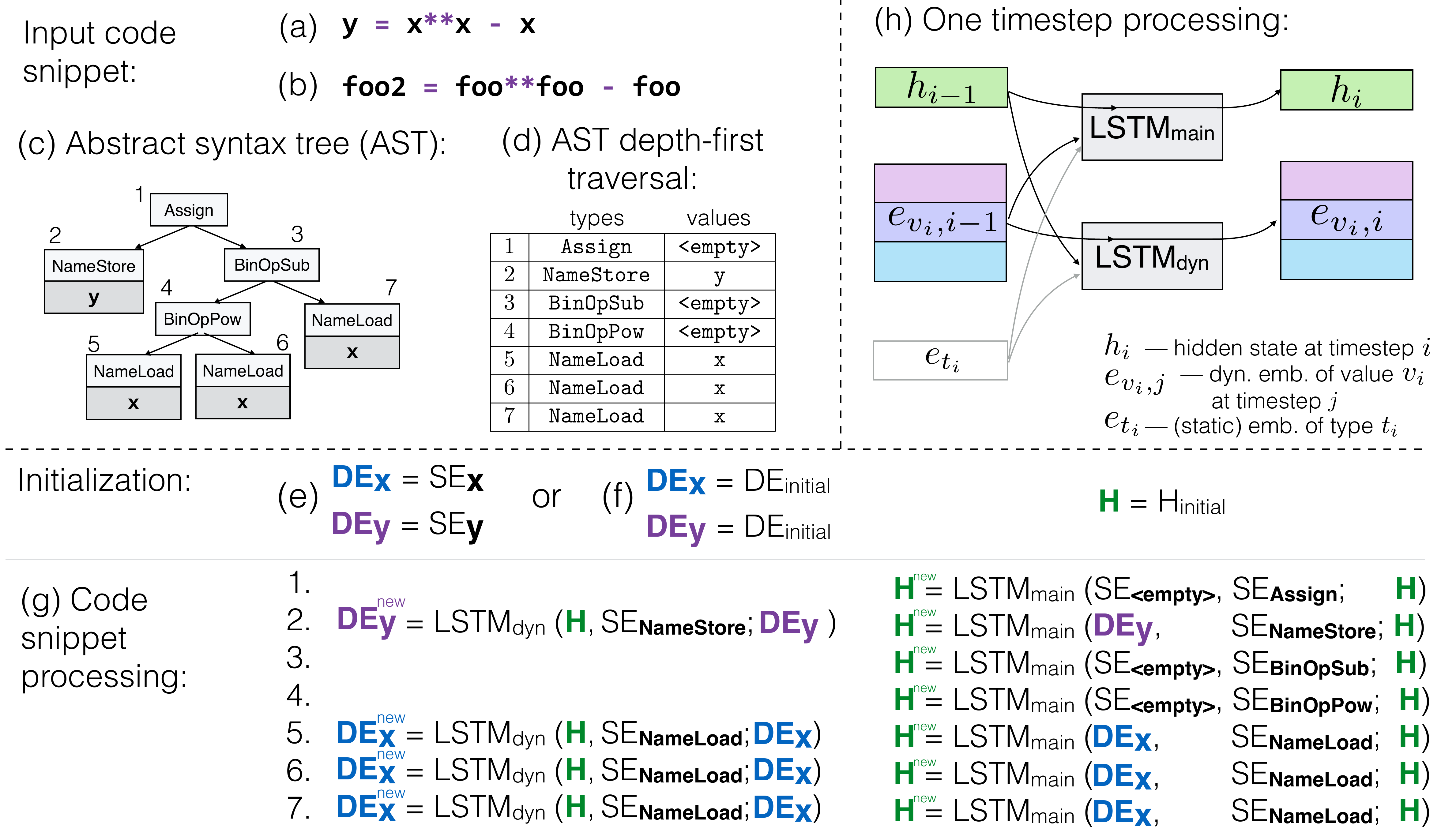} 
        \caption{The overview of the proposed approach. (a) and (b): two variants of the input code snippet, variant (a) is used in other illustration blocks; (c) abstract syntax tree (AST); (d) AST converted to a sequence that will be passed to the RNN; (e) the static-embedding-based initialization of dynamic embeddings; (f) the constant initialization of dynamic embeddings; (h) the scheme of updating dynamic embeddings and hidden states; (g) the scheme of one timestep processing. SE: static embedding, DE: dynamic embedding.}
        \label{fig:illustration}
\end{figure*}

Let us take a closer look at the invariance property. In Fig.~\ref{fig:illustration}  (a) and (b), a code snippet implementing a simple mathematical calculation is presented with two different variable naming schemes. Both code snippets implement the same algorithm, i.\,e.\,are equivalent in the ``program semantics'' space, but have different text representations. Classic NLP approach implies using the embedding layer as the first layer in the network where learnable embeddings encode the global semantics of the input tokens. 
In example (a), the embeddings of variables \verb|x| and \verb|y| make sense, as these variables are usually used in mathematical calculations, but in example (b), the embeddings of variables \verb|foo| and \verb|foo2| do not reflect any semantics.
Moreover, even in case (a), the semantics of identifier \verb|y| reflected by its embedding are too broad, i.\,e.\,this identifier could be used in a lot of different calculations, while variable \verb|y| has a much more specific role in the program, i.\,e.\,storing the result of the particular function. The key idea of this work is that \textit{the embedding of a variable in the program should reflect the variable's particular role in this program and not only its name}. The name of the variable may act as the secondary source of information about the variable's role, but the main source of this information is the program itself, i.\,e.\,the contexts the variable is used in.

We develop the recurrent mechanism called \textit{dynamic embeddings} that captures the representations of the variables in the program based on the contexts in which these variables are used. Being initialized before a program processing, the dynamic embedding of a variable is updated each time the variable has been used in the program, see the scheme in Fig.~\ref{fig:illustration}(g). We test the dynamic embedding approach in two settings: the standard setting with the full data, and the anonymized setting, when all variable names are replaced with unique placeholders \verb|var1|, \verb|var2|, \verb|var3| etc. 
In the full data setting, we initialize the dynamic embeddings with standard embeddings, see Fig.~\ref{fig:illustration}(e), to implement  the idea of the variable name being a secondary source of information about the variable's semantics.
In the anonymized setting, we initialize the dynamic embeddings using a constant initial embedding, the same for all identifiers, see Fig.~\ref{fig:illustration}(f). In this setting, the variable names are not used at all, and the model detects the role of the variable purely based on the contexts in which the variable is used in the program.
Although being less practically oriented, the anonymized setting is a conceptually interesting benchmark, as it highlights the capabilities of deep learning architectures to understand the pure program \textit{structure}, that is actually the main goal of SCP, without relying on the \textit{unstructured} textual information contained in variable names. In addition, the anonymized setting could be the case in practice, e.\,g.\,when processing the decompiled or obfuscated code~\cite{disassemble}. 

In the experiments, we show that using the proposed dynamic embeddings significantly outperforms the model that uses the standard embeddings, called static embeddings in our work, in both described settings in two SCP tasks, namely code completion and bug fixing.

To sum up, our contributions are as follows:
\begin{itemize}
    \item We propose the dynamic embeddings for capturing the semantics of the variable names in source code;
    \item To demonstrate the wide practical applicability of the proposed dynamic embeddings, we show that they outperform static embeddings in two different code processing tasks, namely code completion (generative task) and bug fixing (discriminative task), in the full data setting;
    \item We propose the version of the dynamic embeddings approach that does not use variable names at all, and show that it achieves high results in both tasks, sometimes even outperforming the standard model trained on full data (with variable names present in the data).
\end{itemize}
Our source code is available at \url{https://github.com/nadiinchi/dynamic_embeddings}.

\section{Related Work}
The possibility of improving deep learning models of source code by taking into account the invariance property of variable names has been superficially discussed in the literature. \citet{student_code} replace variables with their types, while \citet{deepfix} and \citet{commitgen} use the static embeddings for anonymized variables. However, the existing SCP work did not consider developing a special architecture that dynamically updates the embeddings of the variables during processing a code snippet.

Our research is also related to the field of processing out-of-vocabulary (OOV) variable names. The commonly used approaches for dealing with OOV variables are using the pointer mechanism~\cite{pointer} or replacing OOV variables with their types~\cite{deepcom}. As we show in our work, both methods may be successfully combined with the proposed dynamic embeddings.

In the context of NLP, \citet{kobayashi} use a similar model with dynamic embeddings to process OOV and anonymized named entities in natural text. In contrast to their approach, we apply dynamic embeddings to the whole vocabulary of variable names, and incorporate dynamic embeddings into the model that relies on the syntactic structure of code. This results in more meaningful dynamic embeddings. 

\section{Proposed method}
We firstly describe what format of the model input we use, i.\,e.\,the procedure of code preprocessing, and then describe our model. At the end of this section, we discuss how we use the proposed model in two code processing tasks.

\subsection{Code preprocessing} To capture the syntactic structure of an input code snippet, we convert it to an abstract syntax tree (AST), see Fig.~\ref{fig:illustration}(c) for the illustration. In order to process the code snippet with an RNN, we need to convert the AST into a sequence. We use the most popular approach that implies traversing the AST in the depth-first order~\cite{pointer}, see Fig.~\ref{fig:illustration}(d). Recent research shows that using the AST traversal may be even more effective than using specific tree-based architectures~\cite{chirkova_troshin}.

Each node in the AST contains a \textit{type}, reflecting the syntactic unit, e.\,g.\,\verb|If| or \verb|NameLoad|. Some nodes also contain \textit{values}, e.\,g.\,a user-defined variable or a constant. We insert the \verb|<EMPTY>| value in the nodes that do not have values so that the input snippet is represented as a sequence of (type, value) pairs: $I = [(t_1, v_1), \dots, (t_L, v_L)]$. Here $L$ denotes the length of the sequence, $t_i \in T$ denotes the type and $v_i \in V$ denotes the value. The size of the type vocabulary $T$ is small and determined by the programming language, while the size of the value vocabulary $V$ may be potentially large, as it contains a lot of user-defined identifiers. Given sequence $I$, the RNN outputs a sequence of hidden states $[h_1, \dots, h_L]$, $h_i \in \mathbb{R}^{d_{\mathrm{hid}}}$, $i=1, \dots, L$. These hidden states are used to output the task-specific prediction as described in Section~\ref{sec:prediction}.

\subsection{Dynamic embeddings}
\label{sec:dynemb}
We use the standard \textit{baseline} recurrent architecture that initializes the hidden state with a learnable predefined vector $h_{\mathrm{init}} \in \mathbb{R}^{h_{\mathrm{hid}}}$: $h_0=h_{init}$, and then updates the hidden state at each timestep $i=1, \dots, L$:
\[
h_{i} = \mathrm{LSTM}_{\mathrm{main}} (e_{v_i}, e_{t_i}; h_{i-1}).
\]
Here, $e_{v_i} \in \mathbb{R}^{d_{\mathrm{val}}}$ and $e_{t_i} \in \mathbb{R}^{d_{\mathrm{type}}}$ denote the embeddings of the value and the type correspondingly. Without loss of generality, we use the Long Short-Term Memory recurrent unit (LSTM)~\cite{lstm}. In this work, we replace value embeddings $e_{v_i}$ with \textit{dynamic} embeddings described below.

\paragraph{Dynamic embeddings.} The general idea of dynamic embeddings is that the variable's embedding is updated in the RNN-like manner after each occurrence of the variable. We first describe the updating procedure and then discuss the initialization strategy. 
Since the dynamic embeddings change over timesteps, we use notation $e_{v, i}$ for the dynamic embeddings of value $v$ at timestep $i$. For example, for the value located at the $i$-th position in the input sequence, $v_i$, the dynamic embedding after processing the $i$-th step is denoted as $e_{v_i, i}$, and its previous state is denoted as $e_{v_i, i-1}$.
At each timestep $i=1,\dots, L$, we update the dynamic embedding $e_{v_i, i}$ of the current value $v_i$ and hidden state $h_{i}$ using two LSTMs:
\begin{equation}
\label{f1}
    e_{v_i, i} = \mathrm{LSTM}_{\mathrm{dyn}} (h_{i-1}, e_{t_i}; e_{ v_i, i-1})
\end{equation}
\begin{equation}
\label{f2}
    e_{v, i} = e_{v, i-1}, \quad v \ne v_i
\end{equation}
\begin{equation}
\label{f3}
    h_{i} = \mathrm{LSTM}_{\mathrm{main}} (e_{v_i, i-1}, e_{t_i}; h_{i-1})
\end{equation}
An illustration of this update procedure is given in Fig.~\ref{fig:illustration}(h), and the example scheme of processing a code snippet is given in Fig.~\ref{fig:illustration}(g).
$\mathrm{LSTM}_{\mathrm{main}}$ implements the recurrence over the hidden state, while $\mathrm{LSTM}_{\mathrm{dyn}}$ implements the recurrence over dynamic embeddings, and the same $\mathrm{LSTM}_{\mathrm{dyn}}$ is used to update the dynamic embeddings of different values at different timesteps. We note that at timestep $i$, the dynamic embedding of only current value $v_i$ is updated, while the dynamic embeddings of other values do not change, as stated in Eq.~\eqref{f2}.

In practice, several dummy values, e.\,g. \verb|<EMPTY>|, \verb|<UNK>| and \verb|<EOF>|, do not change their roles 
in different sequences. We 
use static embeddings for these values.

\paragraph{Initializing dynamic embeddings.} The most reasonable strategy for initializing the dynamic embeddings is to use static embeddings: $e_{v, 0}=e_{v}$ where $e_v$ are the learnable embedding vectors of all values $v$ in vocabulary $V$. In this case, the model utilizes all the available information about the variable: the variable's name introduced by the programmer that is supposed to somehow reflect the mission of the variable, and the contexts in which the variable occurs (captured by hidden states). In other words, the model firstly ``understands'' the loose role of the variable from its name and then ``finetunes'' this understanding, while learning more about what the variable is used for. 

Another possible strategy is to ignore all the variable names and initialize all dynamic embeddings with a constant embedding: $e_{v, 0} = e_{\mathrm{init}}$, $e_{\mathrm{init}} \in \mathbb{R}^{d_{\mathrm{val}}}$, $v \in V$. Although the initial embeddings of all values are the same, they will be updated differently, as different values occur in different locations in the program, and the dynamic embeddings will characterize these locations. Interestingly, the described strategy ensures that if we rename all the variables in the program, the output of the RNN does not change. Such a behaviour is consistent with the variable invariance property: renaming all user-defined variables does not change the underlying algorithm. The common sense is that the architecture for processing source code should be consistent with the variable invariance property, and dynamic embeddings with the constant initial embedding fulfill this conceptual requirement. On the other hand, commonly-used static embeddings are not consistent with the invariance property, i.\,e.\,renaming variables scheme results in using different embeddings and changes the predictions of the RNN.

As will be shown below, in practice, using both sources of information, namely variable names and variable occurrences in the program, performs better than relying on only one source of information. In other words, dynamic embeddings with static embedding initialization outperform both static embeddings in the full data setting (relying only on variable names) and dynamic embeddings with constant initialization (relying only on variable occurrences).

\subsection{Task-specific prediction}
\label{sec:prediction}
We test the proposed dynamic embeddings in two SCP tasks: code completion and variable misuse prediction. Below, we describe how we make predictions in these tasks, using the output $[h_1, \dots, h_L]$ of the RNN.

\paragraph{Code completion.}
In code completion, the task is to predict the next type-value pair $(t_{i+1}, v_{i+1})$ given prefix $[(t_1, v_1), \dots, (t_{1}, v_{i})]$ at each timestep $i=1,\dots,L$. In our work, we focus on value prediction, as type prediction is a simple task usually solved with high quality in practice~\cite{pointer}. We rely on the setup and the architecture of~\citet{pointer}.

To predict the next value $v_{i+1}$ based on $[h_1, \dots, h_{i}]$, we firstly apply the standard attention mechanism~\cite{attention}, obtaining the context vector $c_{i}=\sum_{j=1}^{i}\alpha_j h_j$, $\alpha_j$ denote attention weights, and then combine all available representations using a fully-connected layer:
\[
\hat h_{i} = W^{1} h_{i} + W^2 c_{i} + W^3 h_{\mathrm{parent}},
\]
where $h_{\mathrm{parent}}$ is the hidden state of the parent node. 
For computing the logit $y_{v, i} \in \mathbb{R}$ of each value $v$, we reuse the dynamic embeddings $e_{v, i}$  of the input layer, as well as the static embeddings of several dummy values: 
$y_{v, i} = e_{v, i}^T \hat h_{i}$,
and apply Softmax 
on top of $y_{v, i}$ to predict the probability distribution $P_i^{vals} \in \mathbb{R}^{|V|}$ over next value $v$.
Finally, we use the pointer mechanism to improve the prediction of rare values. We reuse attention scores $[\alpha_1, \dots, \alpha_{i}]$, $\sum_{j=1}^{i}\alpha_{j} = 1$, $\alpha_j \geqslant 0$ as a distribution over previous positions $P_i^{pos} \in \mathbb{R}^{i}$, and use switcher $s = \sigma (w^{swit, 1} h_{i} + w^{swit, 2} c_{i}) \in (0, 1) $ to gather two distributions into one: $R_i = [s P_i^{vals}, (1-s) P_i^{pos}]$. To make the prediction, we select the largest element of vector $R_i$; if it corresponds to the value from the vocabulary, we output this value, if it corresponds to the position, we copy the value from that position.
To train the model, we optimize the cross-entropy loss, using as ground truth the values in the vocabulary for in-vocabulary values and the last occurrence of the value (if any) for out-of-vocabulary values.

\paragraph{Variable misuse prediction.}
The variable misuse task implies outputting two pointers: the first one points to the location $i$ in which the wrong value $v_i$ is used and the second one points to the location $j$ that can be used to repair the bug by copying its value $v_j$. If there is no bug, the first pointer selects a special no-bug location. In this task, we rely on the approach of~\citet{vasic2018neural} and its implementation of~\citet{hellendoorn}. In addition, we change the format of the model input and use the depth-first AST traversal~\cite{pointer}. 

We use the bidirectional LSTM, with each of the two LSTMs being equipped with its own dynamic embeddings. As a result, we have two sequences of hidden states: $[h^{\mathrm{fw}}_1, \dots, h^{\mathrm{fw}}_L]$ and $[h^{\mathrm{bw}}_1, \dots, h^{\mathrm{bw}}_L]$. To make the prediction, we firstly combine two representations using a fully-connected layer:
\[
h_i = tanh(W^1 h^{\mathrm{fw}}_i + W^2 h^{\mathrm{bw}}_i), ~~ i=1, \dots, L
\]
and then use two other fully-connected layers to obtain logits $y^{\mathrm{bug}}_i \in \mathbb{R}$ and $y^{\mathrm{fix}}_i \in \mathbb{R}$ of each position $i$: $y^{\mathrm{bug}}_i = (w^{\mathrm{bug}})^T h_i$, $y^{\mathrm{fix}}_i = (w^{\mathrm{fix}})^T h_i$. Finally, we apply Softmax over $[y^{\mathrm{bug}}_1, \dots, y^{\mathrm{bug}}_L, y^{\mathrm{nobug}}]$ and over $[y^{\mathrm{fix}}_i]_{i=1}^L$ to obtain two distributions over positions. Here, learnable $y^{\mathrm{nobug}} \in \mathbb{R}$ corresponds to a no-bug position. The model is trained using the cross-entropy loss.

\section{Experimental setup}
\label{sec:exp_setup}

\paragraph{Data and preprocessing.} We conduct experiments on Python150k~\cite{python150k} and JavaScript150k~\cite{javascript150k} datasets. Both datasets are commonly used in SCP and were obtained by downloading repositories from GitHub. However, the train~/~test split released by the authors of the dataset does not follow the best practices of splitting data in SCP~\cite{allamanis,sum_naacl}, so we use another train~/~test split released by~\citet{chirkova_troshin}. This split is based on the repositories, i.\,e.\,all files from one repository go either to train or test, and was deduplicated using the tools provided by~\citet{allamanis}, i.\,e.\, code files in the test set that are duplicated in the train set were removed; this is a common case in source code downloaded from GitHub. In addition, the Python dataset includes only redistributable code~\cite{redistr}.
Splitting by repository and deduplication are highly important in SCP to avoid a percentage of testing accuracy being provided by the examples the model saw during training.
With the described new split, the results in our tables are not directly comparable to the results reported in other works. To validate our implementation, we compared the quality of baseline models trained in our implementation with the quality reported in the papers describing these baselines, and observed that the numbers are close to each other (see details in Section~\ref{sec:reproduce}). 

For the code completion task, we use the entire code files as training objects, filtering out exceptionally long files, i.\,e.\,files longer than $3\cdot10^4$ characters. The resulting training~/~testing set consists of  76K~/~39K files for Python and of 69K~/~41K for JavaScript. The mean length of the code files in 567~/~669 AST nodes for Python~/~JavaScript.

For the variable misuse task, we select all top-level functions, including functions inside classes from all files, and filter out functions longer than 250 AST nodes, and functions with fewer than three positions containing user-defined variables or less than three distinct user-defined variables. The resulting training~/~testing set consists of 417K~/~231K functions for Python and 202K~/~108K functions for JavaScript. One function may occur in the dataset up to 6 times: 3 times with a synthetically generated bug and 3 times without bug. The buggy examples are generated synthetically by choosing random bug and positions from positions containing user-defined variables. The described strategy for injecting synthetic bugs is the same as in~\cite{hellendoorn}.

In both tasks, the size of the node type vocabulary is 330~/~44 for Python~/~JavaScript, the vocabulary of node values is limited to 50K of the most frequent values.

\paragraph{Metrics.}
Following~\citet{pointer}, we use accuracy to measure model quality in the code completion task, counting all predictions of \,\verb|<UNK>| as wrong.  Following~\citet{hellendoorn},  to measure the quality in the variable misuse task, we use the joint localization and repair accuracy (what portion of buggy values is correctly located and fixed). 

\paragraph{Details.} In all our models, node type embeddings have 300 units, node value embeddings have 1200 units (for static embeddings), and the one-layer LSTM's hidden state has 1500 units. The described model size matches the configuration of the model of~\citet{pointer}. The proposed dynamic embeddings of values have 500 units in all experiments to show that they outperform the static embeddings with much less dimension.
In the code completion task, we split the input AST traversals into the chunks, each chunk has the length of 50 AST nodes, and apply attention and pointer only over the last 50 positions. In the variable misuse task, we pass the entire function's AST traversal to the model.

In code completion~/~variable misuse tasks, we train all models for 10 epochs with AdamW~\cite{adamw}~/~Adam~\cite{adam} with an initial learning rate of 0.001~/~0.0001, a learning rate decay of 0.6 after each epoch, a batch size of 128~/~32, and using weight decay of 0.01~/~0. We also use early stopping for the variable misuse task. For code completion, all hyperparameters are the same as in~\cite{pointer}. We tuned hyperparameters to achieve convergence on the training set, for variable misuse. We use the same hyperparameters for static and dynamic embedding models. Both datasets are large, which helps to avoid overfitting, thus regularization is not needed.

\begin{figure}[t!]
    \centering
        \centering
        \includegraphics[width=0.75\linewidth]{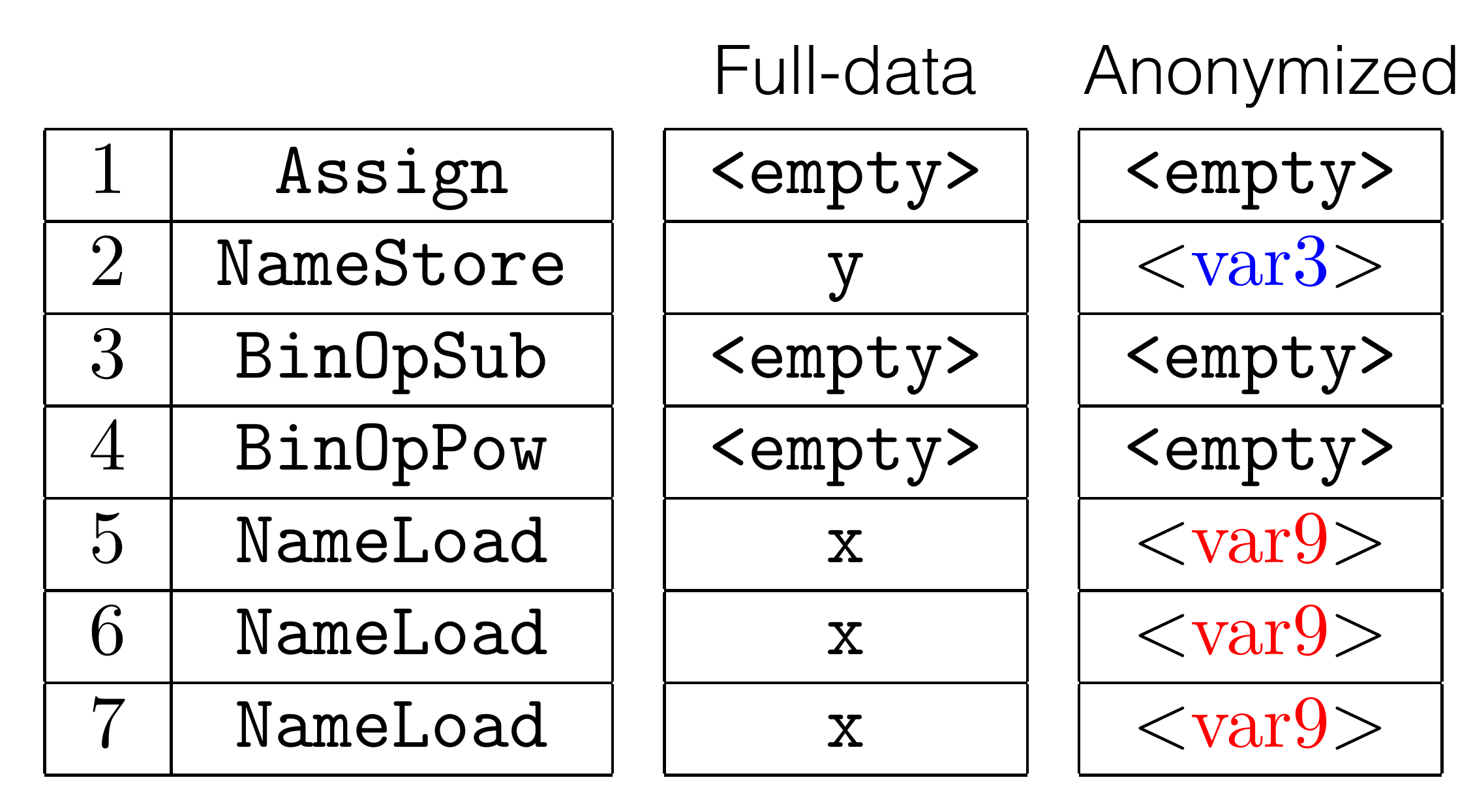}
         \caption{The visualisation of the model input in two settings considered in the paper: full data setting and anonymized setting, for the example code snippet from Figure~\ref{fig:illustration}(a). The leftmost column represents types (the same for both settings), two other columns visualize values for two settings.
        }
        \label{fig:an_illustration}
\end{figure}

\section{Experiments}

\begin{table*}[ht!]
\normalsize
\centering
\begin{tabular}{l|ccc|ccc}
\hline
 & \multicolumn{3}{c|}{\textbf{PY}} &  \multicolumn{3}{c}{\textbf{JS }} \\
\textbf{Model} &  \textbf{LSTM} & \textbf{LSTM+at} & 
\textbf{LSTM+pt} & \textbf{LSTM} & \textbf{LSTM+at} & 
\textbf{LSTM+pt}\\
\hline
Stat.\,emb. (an. data) & 55.76 &	59.74 &	60.28 &   51.80 & 56.26 & 57.67\\
Dyn.\,emb. (an. data) & \textbf{66.35} & \textbf{66.79} & \textbf{66.90} &   \textbf{61.69} & \textbf{62.86} & \textbf{62.85}\\
\hline
Stat. emb. (full data) & 61.62 & 63.73 & 64.69 &   62.03 & 64.28 & 65.05\\ \hline
\end{tabular}
\caption{\textit{Anonymized} setting, code completion task, accuracy (\%) of the proposed dynamic embedding model and the baseline static embedding  model on Python150k (Py) and JavaScript150k (JS) datasets. All standard deviations over three runs are less than 0.05\%.
The last row represents the conventionally used model trained on the full data~\citep{pointer} (this model uses more information during training than the models in the first three rows). 
Columns list the three variants of the base architecture: LSTM, attentional LSTM (LSTM+at), and attentional LSTM with pointer (LSTM+pt).
}\label{tab:python}
\end{table*}

\subsection{Anonymized setting}
\label{sec:anonym_setting}
We firstly test the proposed dynamic embeddings in the setting without using the user-defined variable names, stored in node values. Directly omitting values results in losing much information, this can be seen as replacing \textit{all} the variables in a code snippet with the \textit{same} variable \verb|var|. To save the information about whether two AST nodes store the same value or not, we \textit{anonymize} values, i.\,e.\,we map the set of all node values in the program (except dummy values, e.\,g. \verb|<EMPTY>|) to the 
random subset of anonymized values \verb|var1...varK|, $K$ is a size of the anonymized value vocabulary, we use $K=1000$. For example, code snippet \verb|sum = sum + lst[i]| may be transformed into \verb|var3 = var3 + var8[var1]|, and \verb|stat = [sum / n; sum]| --- into \verb|var1 = [var5 / var2; var5]|. All occurrences of the same value in the program, e.\,g.\,\verb|sum|, are replaced with one anonymized value, but value \verb|sum| may be replaced with different anonymized values in different programs. Fig.~\ref{fig:an_illustration} visualizes how the anonymization is applied to AST.
Although being not so practically oriented, the anonymized setting highlights the capabilities of the deep learning models to capture \textit{pure} syntactic information from the AST, without relying on the unstructured text information laid in variable names. In our opinion, this setting should become a must for the future testing of syntax-based SCP models, and the proposed dynamic embeddings could be used as a first layer in such models to capture an equal-not-equal relationship between values.

In the described \textit{anonymized} setting, we compare the proposed dynamic embeddings (constant initialization) with the
static embeddings, i.\,e.\,learning the static embeddings of \verb|var1..varK|.

\begin{table}[t!]
\normalsize
\centering
\begin{tabular}{l|c|c}
\hline
\textbf{Model} & \textbf{PY} &
\textbf{JS}\\ \hline
Stat. emb. (an. data) & 25.17 & 13.16  \\
Dyn. emb. (an. data) & \textbf{63.64} & \textbf{53.53} \\ \hline
Stat. emb. (full data) & 54.78 & 35.06 \\ \hline
\end{tabular}
\caption{\textit{Anonymized} setting, variable misuse task, joint accuracy (\%) of the proposed dynamic embedding model and the baseline static embedding model on Python150k (Py) and JavaScript150k (JS) datasets. All standard deviations over three runs are less than 0.1\%.
The last row represents the conventionally used model trained on the full data~\cite{vasic2018neural} (this model uses more information during training than the models in the first three rows). 
}
\label{tab:vm_an}
\end{table}

\paragraph{Results for the code completion task.} In the code completion task, 
we consider three variants of the architecture: plain LSTM, and attentional LSTM with and without pointer. We note that our goal is to compare the dynamic embeddings with the baseline in three setups, i.\,e.\,using three base architectures. We do not pursue the goal of comparing base architectures. Table~\ref{tab:python} lists the results.

For all base architectures, the proposed dynamic embeddings outperform static embeddings by a large margin. We note that the number of parameters in both architectures is approximately the same. In the first two setups, with plain and attentional LSTMs, the models can only predict values by generating them from the vocabulary (no pointer), relying on the input and output embeddings of the values. In these setups, the difference between static and dynamic embeddings is large,
indicating that dynamic embeddings capture the semantics of the variables significantly better.
In the setup with pointer LSTM as a base architecture, the static embeddings  win back some percent of correct predictions by relying on the pointer mechanism. Still, the gap between them and dynamic embeddings is large. The portion of correct predictions made using the pointer is 25\% for static embeddings and only 0.01\% for dynamic embeddings. This shows that dynamic embeddings actually replace the pointer mechanism, performing better. This also explains why the difference in quality of dynamic embeddings between attentional LSTM and pointer LSTM is very small. 

\begin{table}[t!]
\normalsize
\centering
\begin{tabular}{l|cc|cc}
\hline
\textbf{Model} & \multicolumn{2}{c|}{\textbf{Code compl.}} & \multicolumn{2}{c}{\textbf{ Var. misuse}} \\
 (full data) & \textbf{PY} & \textbf{JS} & \textbf{PY} & \textbf{JS} \\
\hline
Stat. emb. & 64.69 &   65.05 &    54.78 & 35.06 \\ 
Dyn. emb. & \textbf{68.61} &   \textbf{65.67} &   \textbf{68.59}   & \textbf{53.74} \\
\hline
\end{tabular}
\caption{\textit{Full data} setting, two tasks, Python150k (Py) and JavaScript150k (JS) datasets. Accuracy (\%) of LSTM with pointer (code completion), joint accuracy (\%) of BiLSTM (variable misuse). All standard deviations are less than 0.05\% for code completion and 0.1\% for the variable misuse task. Comparing the conventionally used model (static embeddings) and the proposed dynamic embeddings (static initialization). The conventionally used model is a model of~\citet{pointer} for the code completion task and of~\citet{vasic2018neural} for the variable misuse task. Note that we use custom data split, see details in Sec.~\ref{sec:exp_setup}.}\label{tab:full_data}
\end{table}

Interestingly, on the Python dataset, the model with dynamic embeddings trained in the anonymized setting outperforms the conventionally used static embedding model trained in the full data setting, although the first model uses much less information during training. 
The explanation is that the first model predicts rare values much better than the second model: the accuracy of rare\footnote{By rare values, we mean values outside top-1000 frequent values.} values prediction is 27\% for the first model and 11\% for the second, for the pointer LSTM model. On the contrary, frequent values are easier to predict with static embeddings: the accuracy of predicting frequent values is 53\% for the first model and 57\% for the second model. The total frequencies of rare and frequent values are approximately the same and equal to 25\% (the rest 50\% are \verb|EMPTY| values, they are predicted with similar quality with both models). As a result, when counting accuracy over all values, the first model outperforms the second one. 

However, on the JavaScript dataset, the first model does not outperform the second one. We analysed the example predictions of both models on both datasets and found that in JavaScript, there are a lot of short code snippets commonly used in different projects. This is expected since JavaScript is mostly used for one purpose, web development, while Python is used for a lot of different purposes. As a result, for JavaScript, the total frequency of top-1000 values is 32\% (higher than for Python), while the total frequency of rare values is 22\% (less than for Python). The commonly used code snippets are easy to predict in the full data setting but hard to predict in the anonymized setting: the accuracy of predicting frequent values is only 44\% for the first model and 54\% for the second model. The rare values are still better predicted with dynamic embeddings, but with the gap smaller than for Python: the accuracy of rare values prediction is 23\% for the first model and 17\% for the second one. The gap is smaller since rare values also occur in the commonly used code snippets which improves the performance of the second model on rare values. When counting accuracy over all values, the second model outperforms the first one.

\begin{figure*}[h!]
    \centering
         \includegraphics[width=\linewidth]{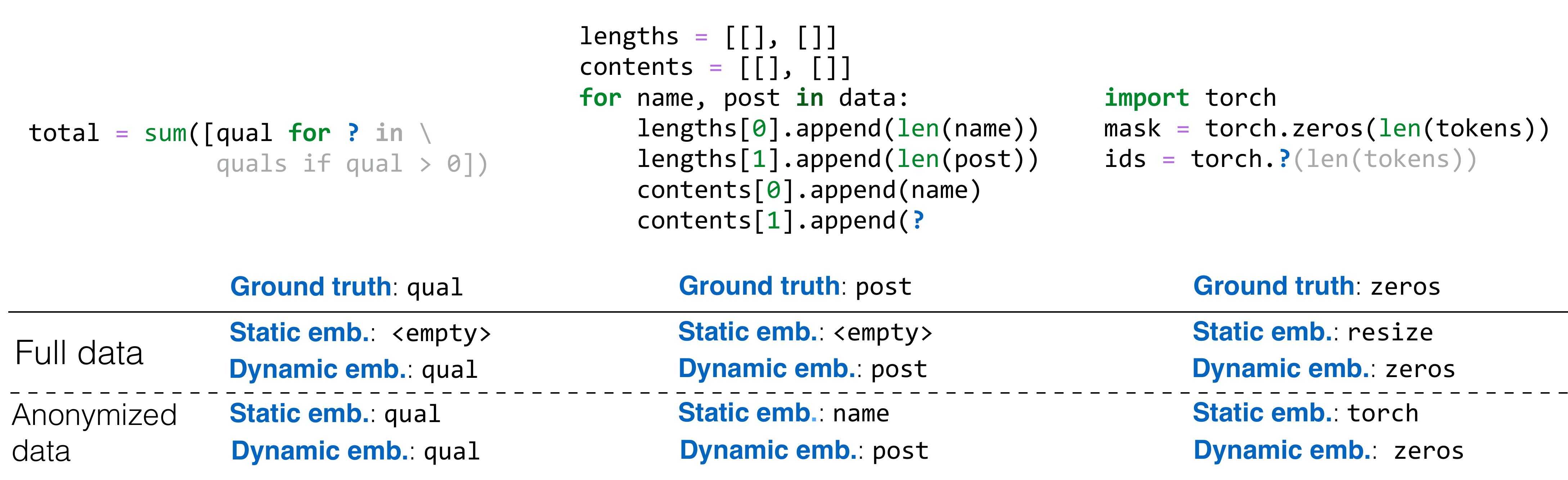} 
        \caption{Example predictions for code completion task on Python language. Row 1: ground truth; rows 2 and 3: model trained in the full data setting; rows 4 and 5: models trained in the anonymized setting (these models observe data in a different way, see Fig.~\ref{fig:an_illustration}). The model predicts one next token based on the prefix and does not see gray-colored code. }
        \label{fig:example}
\end{figure*}

\paragraph{Results for the variable misuse task.} Table~\ref{tab:vm_an} lists the joint accuracies of the proposed model and the baseline in the anonymized setting.
Again the dynamic embeddings outperform static embeddings by a large margin. Moreover, the dynamic embeddings outperform even the commonly used static embedding model trained on the full data, for both datasets. We think the reason is that we use the dynamic embeddings in two layers of bi-directional LSTMs and these bi-directional dynamic embeddings provide a rich representation of the input code snippet.

\subsection{Full data setting}
\label{sec:fulldata_setting}
We now test the proposed dynamic embeddings in the full data setting, i.\,e.\,we compare a commonly used model with static embeddings and the proposed model with dynamic embeddings (static embedding initialization). The initialization of dynamic embeddings was discussed in Sec.~\ref{sec:dynemb}. Both models process the full data (see illustration in Fig.~\ref{fig:an_illustration}.

The results for both tasks are presented in Table~\ref{tab:full_data} and show that the dynamic embeddings outperform the static embedding model in all cases. We note that dynamic embeddings could be easily incorporated into any recurrent SCP architecture. In our experiments we incorporate them into the base models of~\citet{pointer} and \citet{vasic2018neural} and show that the dynamic embeddings significantly improve these base models. We also note that we use the dynamic embeddings of 500 units while static embeddings have 1200 units. The number of parameters in the dynamic embedding layer, 2.6M, is much smaller than that of the main LSTM layer, 13.8M, and two orders smaller than the number of parameters in the embedding layer, 134M (the numbers are given for the code completion task). 

\subsection{Example predictions}
Figure~\ref{fig:example} visualizes the predictions of different models for three example code snippets in Python. We highlighted three scenarios when the dynamic embedding model outperforms the static embedding model in the full data setting: 1) capturing the specific role of the variable, e.\,g.\,variable \verb|qual| indexes sequence in the list comprehension in the left example; 2) associating variables with each other, e.\,g.\, in the central example, variable \verb|name| always goes with \verb|0|, and variable \verb|post| always goes with \verb|1|; 3) repeating variables when they occur in the similar context they have already been used,  e.\,g.\,\verb|zeros| in the right example. In all these examples, the proposed dynamic model makes correct predictions, while the static embedding model makes mistakes, in the full data setting.
In the anonymized setting, all models tend to predict previously used variables, and again the dynamic embedding model captures the described relationships, and the static embedding model tends to simply predict the most frequent previously used variable.

\subsection{Validating the implementation}
\label{sec:reproduce}
In our experiments, we use the setup of~\citet{pointer} in the code completion task and of~\citet{hellendoorn} in the variable misuse task, but with our custom data split, see details in Section~\ref{sec:exp_setup}. To maintain the possibility of comparing our results to these works, we trained the static embedding models in the full data setting, with the commonly used train~/~test splits of Python150k and JavaScript150k datasets. 
For code completion with vocabulary size 50K and pointer network, using exactly the same setup as in~\cite{pointer}, we achieved accuracy of 69.39\%~/~80.92\%, while the paper reports 71\%~/~81.0\% for Python~/~JavaScript: the results are close to each other. In the variable misuse task, we achieved joint accuracy of 50.2\% while \citet{hellendoorn} report 44.4\% (Python, JavaScript was not reported in the paper). Our result is higher, since we use 1500 hidden units while \citet{hellendoorn} uses 256 hidden units. In addition, we use different preprocessing and different synthetically generated bugs.

\section{Conclusion}
In this work, we presented dynamic embeddings, a new approach for capturing the semantics of the variables in code processing tasks. The proposed approach could be used in any recurrent architecture. We incorporated dynamic embeddings in the RNN-based models in two tasks, namely code completion and variable misuse detection, and showed that using the proposed dynamic embeddings improves quality in both full data setting and the anonymized setting, when all user-defined identifiers are removed from the data. 

\section*{Acknowledgments}
I would like to thank Sergey Troshin, Irina Saparina, and the anonymous reviewers for the valuable feedback. The results for the anonymized setting presented in section~\ref{sec:anonym_setting} were supported by Samsung Research, Samsung Electronics. The results for the full data setting presented in section~\ref{sec:fulldata_setting} were supported by the Russian Science Foundation grant \textnumero 19-71-30020. The research was supported in part through the computational resources of HPC facilities at NRU HSE.

\bibliography{anthology,custom}
\bibliographystyle{acl_natbib}

\end{document}